\documentclass[prb,twocolumn]{revtex4-2}

\usepackage{}
\usepackage{epsfig}
\usepackage{graphicx}
\usepackage{amsfonts,amssymb}
\usepackage{amsmath}
\usepackage{color}
\usepackage {times}

\definecolor{refcolor}{rgb}{1.0,0.0,0.0}

\begin{document}

\title{Crossover from self-trapped bound states to perturbative scattering\\
 in the Heisenberg-Kondo lattice model}

\author{Tanmoy Mondal$^1$ and Pinaki Majumdar$^2$}

\affiliation{
$~^1$ Harish-Chandra Research Institute (A CI of Homi Bhabha National Institute), 
Chhatnag Road, Jhusi, Allahabad 211019\\
$~^2$ School of Arts and Sciences, Ahmedabad University, Navrangpura, Ahmedabad,
India 380009}
\date{\today}

\begin{abstract}
We map out the complete transport phase diagram of the ferromagnetic 
Heisenberg-Kondo lattice model in two dimensions. The model involves tight-binding electrons, with hopping $t$, coupled to classical spins with a 
coupling $J'$, while the spins have a nearest neighbour coupling $J$ between 
them.  We work with a fixed, small, $J/t$, and study the temperature dependence 
of resistivity for varying electron density $n$, and coupling $J'/t$. Our 
magnetic configurations are generated by exact diagonalisation based Langevin 
dynamics, while the conductivity is computed using the Kubo formula on exact 
eigenstates.  We work on lattices of size $20 \times 20$ and can access 
electron density down to $n \sim 0.01$. The electron system remains homogeneous 
either when the mean density is large or when the coupling $J'$ is small. In 
these situations, the resistivity $\rho(T)$ displays a monotonic increase with 
temperature, and can be understood within a perturbative framework. However, 
at very low density, $n \lesssim 0.05$, strong coupling, $J'/t \gtrsim 1$, 
and for $T \sim T_c$, the electrons can locally polarise the magnetic state, 
create a trapping potential, and form a bound state in it. The resistivity 
associated with this polaronic phase is distinctly non-monotonic, with a 
peak near $T_c$.  We establish the boundary that separates the many body 
polaronic window from traditional scattering and extract a universal form 
for the resistivity in the scattering regime. We suggest the origin of the 
`excess resistivity' in the polaronic regime in terms of an increasing 
fraction of localised states as the temperature tends to $T_c$. This pushes
the mobility edge towards the chemical potential $\mu$ and results in 
enhanced scattering of momentum states near $k_F$. While our specific results 
are in two dimensions, the phenomenology that we uncover should be valid even 
in three dimensions.  
\end{abstract}

\maketitle
~
\section{Introduction}

For non-disordered electrons coupled to a spin system
that transits from a 
ferromagnet at $T=0$ to an essentially uncorrelated paramagnet
at $T \gg T_c$ there are two aspects of the resistivity 
$\rho(T)$ that are
obvious. First, the resistivity vanishes as the temperature
$T \rightarrow 0$, and second, the resistivity saturates when 
$T \gg T_c$. The questions are in the details,
for example about 
(i)~the scattering rate when $T \ll T_c$, 
(ii)~the saturation value of the resistivity when 
$T \gg T_c$, and, most
importantly, (iii)~the nature of $\rho(T)$ when $T \sim T_c$.
While (i) and (ii) are relatively straightforward, since the
magnetic state is known, (iii) is the most interesting 
since the magnetic state has non trivial spatial correlations,
and, additionally,  
at strong electron-spin coupling and low electron density
the electrons can self-trap into magnetic polarons
\cite{ref_EuS,ref_EuS_res,ref_EuGdSe,ref_EuO_res,ref_EuO,ref_exp_raman,ref_exp_muon,ref_exp_carrier_density,ref_exp_resistivity,ref_exp_Mn_pyrochlore,ref_exp_LaCaMn,ref_exp_stm,ref_exp_GdSi,ref_exp_GdN_1,ref_exp_many,ref_exp2,ref_exp3,ref_exp4}.

The scattering regime has long been explored, most importantly
by Fisher and Langer \cite{ref_theo_res_fisher}, 
and the polaronic regime has seen some
work in the last two decades
\cite{ref_theo_other_montecarlo,ref_theory_1polaron,
ref_theo_other_variation,ref_theo_other_dmft,
ref_theory_magneto_resistance,ref_theory_polaron_hopping,
ref_theory_resistivity_phonon,ref_yu2005,ref_theory_ising,
ref_EuB6}. This paper aims to establish
a comprehensive picture connecting the two. 
The difficulty of the problem arises from two sources. 
The first relates to the magnetic state.
In the presence of electrons, the magnetic correlation between the spins is no longer governed only by the Heisenberg
superexchange but also involves electron mediated interactions.
This is well understood in the weak coupling RKKY limit
\cite{ref_theory_magneto_resistance,ref_rkky}
but does not have a simple analytic form at stronger coupling. 
So, even if we knew the magnetic correlations and $T_c$ of the 
Heisenberg ferromagnet, we do not generally know the magnetic 
state in the presence of electrons. 
The second problem is related to electronic properties in
the equilibrium magnetic configurations. If $J'$ were small,
we can just compute resistivity, etc, in low order
perturbation theory. When $J'/t \gtrsim 1$, however, `naive'
perturbation theory does not work. At high density, 
multiple scattering becomes important, and at low
density, polarons can form. Most materials involve
$J'/t \gtrsim 1$, with metals having electron density
$n \sim O(1)$ \cite{ref_metals1, ref_metals2}
and doped semiconductors having $n \sim
10^{-3} - 10^{-2}$
\cite{ref_EuS,ref_EuO_res,ref_exp_carrier_density,
ref_exp_GdSi,ref_exp_GdN_1}. We need to go well beyond 
a Fisher-Langer scheme.

To generate the magnetic state accurately, retaining the electronic 
effect on the spins, and the thermal fluctuations, we use a Langevin 
dynamics scheme that diagonalises the electron problem at every time 
step \cite{ref_spin_lang,ref_spin_lang1,ref_chern}.
In terms of accessing equilibrium configurations, this 
approach is equivalent to exact diagonalisation based
Monte Carlo \cite{ref_theo_other_montecarlo,ref_ed_monte1,ref_ed_monte2}, 
widely used in manganite like problems.
After equilibration we compute the conductivity by using
the exact electronic eigenstates computed 
on the magnetic backgrounds.
Using a parallelised algorithm that scales as $N^2$, where
$N$ is the system size, we can access system size $\sim 20
\times 20$ within a reasonable time.
The method can handle arbitrary $J'/t$ and temperature, and
can access electron density down to $\sim 0.01$ - necessary
for polaron physics. 

While the crossover from a polaronic regime to a scattering
dominated regime leaves its imprint on several measurables,
e.g, optical conductivity, resistance noise, and muon 
scattering, in this paper, we focus only
on the resistivity since it is the easiest to measure.
Our main results are the following.

1.~Using the electronic inverse participation ratio (IPR) and the
resistivity, we establish a boundary in the $n-J'/t$ plane that 
separates the polaronic window from the scattering dominated 
transport regime. Polarons require a critical coupling $J'
> J'_c(n)$, shown in Fig.1. We also establish the thermal 
window around $T_c$ over which polaronic signatures exist.

2.~Even at low densities relevant for polarons, the temperature 
associated with the emergence of ferromagnetic correlations strongly
depends on $n$ and $J'$, since we are in a regime $J' \gg J$. 
We extract this temperature $T_c(n,J')$, which acts as the
normalising scale in our resistivity.

3.~We observe and explain a rough `universal' form for the 
resistivity $\rho(T,n,J')$ away from the polaronic regime. 
This is exact both
at weak and strong coupling and interpolates in between. 
This allows us to collapse the metallic resistivity
to a form $\rho(T) = \rho_m(T) = \rho_{\infty} f(T/T_c)$,
where $\rho_{\infty}$ is the high $T$ value, with a physically
transparent $n$ and $J'$ dependence, and $f$ is a 
normalised function. 

4.~In the polaronic window $\rho(T)$ strongly deviates from 
the metallic form $\rho_m(T)$ near $T_c$. The overall scale
of the `excess resistivity' near $T_c$, $\delta \rho(T) = 
\rho(T) - \rho_m(T)$, is set by $\rho_{\infty}$. We explain 
it's unusual
temperature dependence, with a $d \rho/dT < 0$ window,  from
the behaviour of the electronic lifetime as the polaronic 
mobility edge approaches the chemical potential $\mu$.

The paper is organised as follows. In the next section, we briefly
outline our computational strategy. Following that, successive 
sections discuss the transport
phase diagram, the magnetisation and $T_c$, the resistivity, 
and the nature of electronic eigenstates. We end with a 
discussion of the scattering rate extracted from the single particle
spectral function, and its correlation with the resistivity.

\section{Model and method}

We consider a square lattice with a spin ${\bf S}_i$,
of unit magnitude, at each site, exchange coupled to the nearest 
neighbours by a coupling $J$. The spins are additionally 
coupled locally to electrons via a coupling $J'$, and the
electrons hop between nearest neighbour sites with an
 amplitude
$t$.
\begin{eqnarray}
H &=& 
~~~~~~
H_{el}~~ + ~~H_{el-sp}~~ + ~~H_{sp} \cr
\cr
&=&
 -t \sum_{\langle ij \rangle}^{\sigma} c^{\dagger}_{i\sigma} c_{j\sigma}
- J'\sum_i {\bf S}_i.{\vec \sigma}_i 
-J \sum_{\langle ij \rangle} {\bf S}_i.{\bf S}_j
\end{eqnarray}
We set $t=1$ and $J/t = 0.1$. 
In 3D, the classical Heisenberg model has a 
$T_c \sim 1.5J$.
Our electron number is $N_{el}$,  lattice size $N= L^2$,
and electron density $n = N_{el}/N$. We vary the density 
in the range $1-12\%$.

In 2D, there is no genuine long-range order at $T \neq 0$, 
but on any finite $L \times L$ 
lattice there is a `correlation temperature' $T_{corr}(L)$ 
where the
magnetisation rises sharply as the temperature is lowered.  
It is known that $T_{corr} \sim 1/log(L)$ \cite{ref_hei} 
for the 2D Heisenberg model.
Our intent is to establish (i)~the variation of the $T_{corr}$ 
scale with $n$ and $J'$, staying at $L=20$, and (ii)~the 
correlation between the magnetic and electronic state.
These can be established, and spatial correlations 
visualised readily, in reasonably large 2D lattices. 
With this in mind, for notational
simplicity we will just denote the correlation 
temperature $T_{corr}(L)$ by $T_c$.

At finite $J'$ and $n$ the effective magnetic model that 
determines the spin configurations is not simply $H_{sp}$ 
but has corrections coming from the electrons. Done exactly, 
the effective magnetic Hamiltonian is:
\begin{equation}
H_{sp}^{eff} = H_{sp} - {1 \over \beta}
log[Tr~e^{-\beta (H_{el} + H_{el-sp})}] 
\end{equation}
where the trace is over electronic states in a given spin
background.
There are several ways to proceed from here. If $J'$ were moderate
we can extract the $O(J'^2)$ correction to $H_{sp}$ in the form
of a RKKY term. At even larger $J'$, one can diagonalise
$H_{el} + H_{el-sp}$ for each spin configuration and add
the configuration dependent electronic free energy to $H_{sp}$ 
Given this principle, one can use either Monte Carlo, or computing
the ``update cost'' via diagonalization (an expensive process)
or use $H_{sp}^{eff}$ to generate an effective torque driving
the
spin dynamics. We use the second approach, detailed below,
in the form of a Langevin equation, where the spins are
subject to a systematic torque arising from $H_{sp}^{eff}$
and a stochastic torque proportional to $k_BT$.
This approach affords access to equilibrium spin 
configurations as well as the spin dynamics.

The Langevin equation governing the
evolution of spins is given by:
\begin{eqnarray}
\frac{d\mathbf{S}_i}{dt} &=& \mathbf{S}_i \times (\mathbf{T}_i
+ \mathbf{h}_i) - \gamma \mathbf{S}_i \times (\mathbf{S}_i
\times \mathbf{T}_i) \cr
\mathbf{T}_i &=& -\frac{\partial H}{\partial \mathbf{S}_i} =
-J' \langle \vec{\sigma}_i \rangle -
J \sum_{\langle j \rangle} \mathbf{S}_j
\cr
\langle h_{i\alpha} \rangle &=& 0,~~
\langle
 h_{i \alpha}(t)h_{j \beta}(t') \rangle = 2\gamma k_B T
 \delta_{ij} \delta_{\alpha \beta} \delta(t-t')
\end{eqnarray}
Here $\mathbf{T}_i$
is the effective torque acting on the spin at the $i$-th
 site, $\gamma$ is a damping constant set to $0.1t$, 
and $\mathbf{h}_i$ is a thermal noise satisfying the 
fluctuation-dissipation theorem.
$\langle \vec{\sigma}_i \rangle$ represents the
expectation of $\vec{\sigma}_i$ taken over the
the instantaneous ground state of the electron
in the spin configuration of the previous time step.

The primary indicators that we extract from our data are:
(i)~The instantaneous electron density:
$ n_i^{\alpha} =\sum_n |\psi^{\alpha}_{i,n}|^2$,
where $\alpha$ is a spin configuration index and 
$\psi_n$ is the eigenvector corresponding to
the eigenvalue $\epsilon_n$, 
obtained by diagonalising the electronic 
part of the Hamiltonian over spin configurations $\{ \alpha\}$. 
The sum runs over the number of occupied states.

(ii)~We calculate the optical conductivity $\sigma ( \omega)$ at 
low frequency to estimate the d.c conductivity. For a given
spin background:
$$
\sigma ( \omega) =  {A \over N}
\sum_{m,n} (f_n  - f_m)
{ {\vert j_{mn} \vert^2} \over {\epsilon_{m} - \epsilon_{n}}}
\delta(\omega - (\epsilon_n - \epsilon_m))
$$
The constant $A= (\pi e^2)/ {\hbar} $. 
The matrix element $j_{mn} = \langle \psi_m \vert j_x \vert \psi_n 
\rangle$, and the current operator
in the tight binding model is
$j_x = i t a_0 e \sum_{i, \sigma} (c^{\dagger}_{{i + x a_0},\sigma}
c_{i, \sigma} - h.c)$. The $\psi_m$ etc are single particle eigenstates,
for a given spin configuration and $\epsilon_m, \epsilon_n$, etc
are the corresponding eigenvalues. The   
$f_m $ etc are Fermi factors.
The d.c. conductivity in a given spin configuration $\alpha$ 
is  $\sigma_{\alpha}$ 
$$
\sigma_{\alpha} = {1 \over {\Delta \omega}} 
\int_0^{\Delta \omega} d \omega \sigma_{\alpha} (\omega)
$$
where $\Delta \omega$ is a small multiple of the average finite
size gap in the spectrum. In our calculations, it is $0.05t$.
The d.c. resistivity is the inverse of the thermally averaged
conductivity:
$ \rho = [{1 \over {N_{\alpha}}} \sum_{\alpha} \sigma_{\alpha}]^{-1}$.
(iii)~The inverse participation ratio (IPR) 
is $ I^{\alpha}_n = \sum_i \vert \psi_{i,n}^{\alpha}\vert^4 $.
The inverse of this is an `area' associated with the eigenstate,
$A_{\alpha} = 1/I_{\alpha}$
(iv)~The density of states is: $ D_{\alpha}(\omega) = 
\sum_n \delta(\omega - \epsilon_n^{\alpha})$.

\section{Transport character near $T_c$}

Polaronic effects show up near $T_c$, where the magnetic
system is most polarisable. Based on the electronic 
wavefunctions that arise on equilibrium magnetic 
configurations we can calculate the degree of localisation, 
which we present later in the paper. However, the simplest 
indicator of polaronic behaviour, as we argue later, is a 
peak in the resistivity near $T_c$. We will analyse this 
peak feature
in the resistivity and correlate it to the spatial physics
later, at the moment, we just show a phase diagram.

In Fig.1(a), the non polaronic regime is 
defined by an essentially homogeneous density and a 
monotonic temperature dependence in the resistivity.
The low density strong coupling region, however,
involves localised electronic wavefunctions near $T_c$
(in a typical equilibrium configuration) and shows
a peak in $\rho(T)$ near $T_c$. The  phase boundary 
is defined by the vanishing of that peak as one moves
out of the polaronic window.

\begin{figure}[b]
\centerline{
\includegraphics[height=5.cm,width=8.5cm]{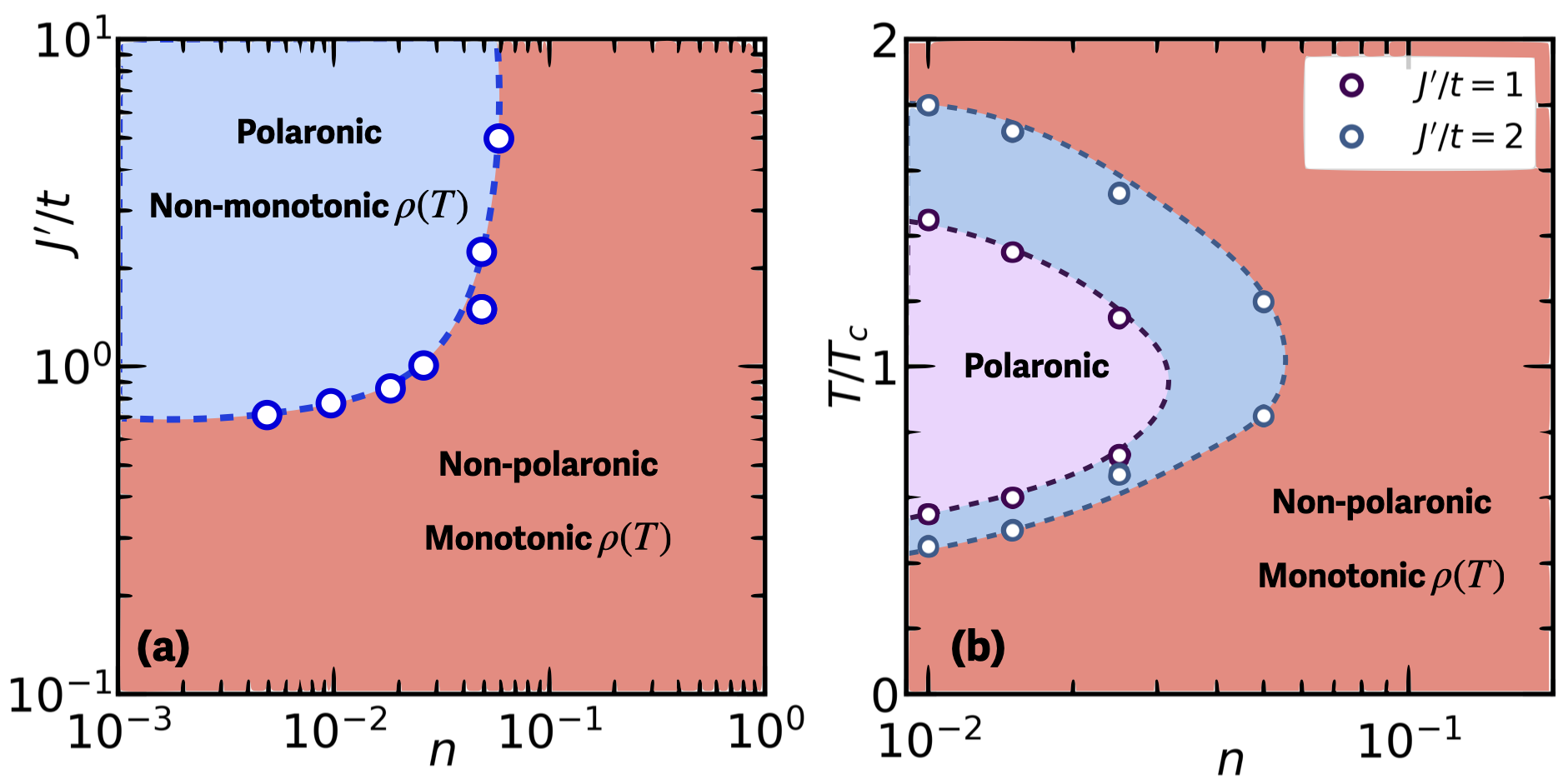}
}
\caption{Transport phase diagram. (a)~The polaronic and
non polaronic windows in the $n-J'$ plane. In the non
polaronic window, the resistivity $\rho(T)$ has a monotonic
rise through $T_c$ to a saturated high $T$ behaviour. In the
polaronic regime, there is a peak in the resistivity around
$T_c$. This peak feature correlates with enhanced localisation
of electronic states around $T_c$. (b)~The temperature window
over which polaronic effects are visible, identified through
the resistivity.
}
\end{figure}
\begin{figure}[t]
\centerline{
\includegraphics[height=7cm,width=8.5cm]{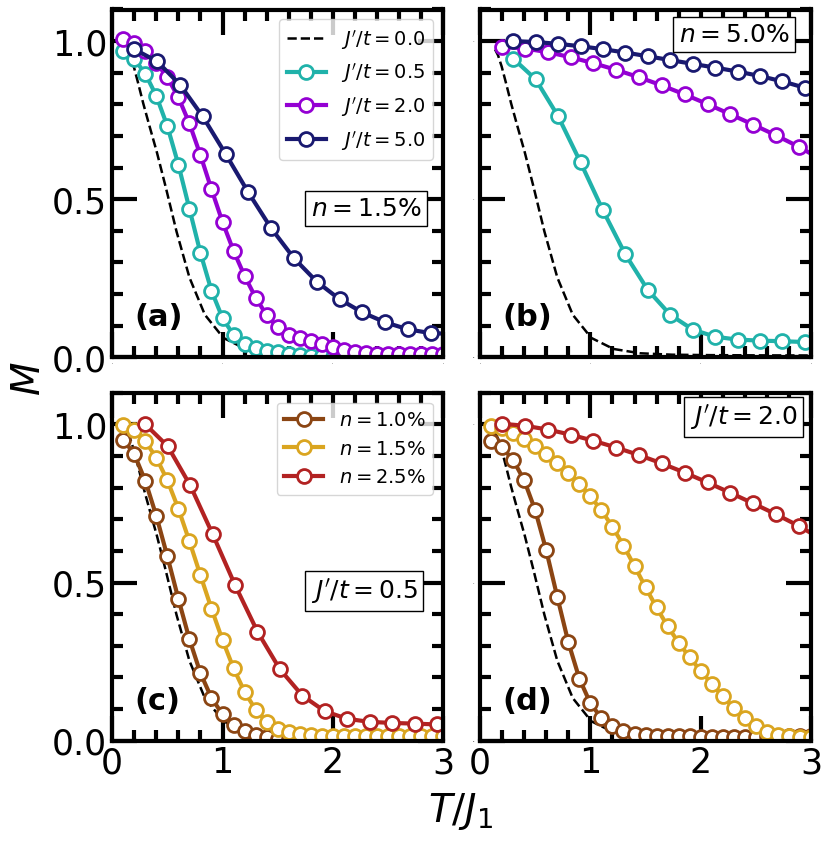}
}
\caption{Temperature dependence of magnetisation for different
$n$ and $J'$. (a)-(b) Variation of $J'$ at low density $n=1.5\%$,
and somewhat larger density, $n=5\%$, respectively. (c)-(d)
Variation of $n$ at $J'/t=0.5$ and $2.0$. The highest
density we have studied in this paper is $n \sim 12\%$, that
data is not shown here to keep the $T$ scale manageable.
}
\end{figure}

The two main features in Fig.1(a) are the existence of 
(i)~a minimum coupling $J'_{min} \sim 0.6t$ below 
which even single polarons do not form, 
and (ii)~a maximum density, 
$n_{max} \sim 0.05$, above which polarons do not exist, 
however large the coupling.
The presence of $n_{max}$ suggests 
a minimum `size' (or area $A_{min}$)
to the polarons, set by $n_{max} A_{min} =1$. Given 
the value of $n_{max}$ this area is $\approx 20$ sites,
suggesting a linear dimension to the polaron $\sim 4$
lattice spacings. 

Fig.1(b) shows the temperature window, normalised by $T_c$,
within which polaronic signatures are seen for a given
$n$ and $J'$. The low temperature state is fully polarised
so electronic textures cannot form, while the $T \gg T_c$
state is entropy dominated, and again polaronic effects
vanish. Between them is the region where $\rho(T)$ shows 
an `excess resistivity', with respect to the expected
background, and the occupied states show localisation.
Fig.1(a)-(b) define the $n-J'-T$ transport 
phase diagram of the model.

\section{Magnetisation and $T_c$}

In the presence of electrons, the magnetic state is no
longer determined purely by the Heisenberg coupling. 
While the ground state continues
to be ferromagnetic upto reasonably high density, 
the effective model describing 
the magnetism takes the following form at weak 
and strong coupling, respectively:
\begin{eqnarray}
H_{eff} &=& -J \sum_{\langle ij \rangle} {\bf S}_i.{\bf S}_j
~+~ J'^2 \sum_{i\neq j} \chi_{ij} {\bf S}_i.{\bf S}_j, 
~~~~~~~~~~~~~ J'/t \lesssim 1 \cr
H_{eff} &=& -J \sum_{\langle ij \rangle} {\bf S}_i.{\bf S}_j 
- t f(n) \sum_{\langle ij \rangle} \sqrt{{ {1 + {\bf S}_i.{\bf S}_j}} 
\over 2}, 
~~~~ J'/t \gg 1
\nonumber
\end{eqnarray}
$f(n)$ is a density dependence that vanishes as $n \rightarrow 0$ and
is maximum at $n=0.5$.
The corrections above are, respectively, the RKKY interaction and
(an approximate form of) the double exchange interaction.
They both enhance the ferromagnetic exchange compared to $J$.
We are working in a regime where $J \ll t$ so the effective 
exchange, at a given $n$, would rise as $J'^2$ from the
Heisenberg limit and saturate to a value $\propto t$ at
large $J'$.

As we have mentioned before, there is no true $T_c$ in 2D but
only a size depedent correlation temperature $T_{corr}(L)$,
which we denote as $T_c$ for convenience.
Fig.2(a)-(b) show the $J'$ dependence of the magnetisation
$m(T)$ at densities $n=1.5\%$ and $n=5\%$, respectively.
The dotted curve is the reference Heisenberg result, and the
temperature is scaled with respect to the Heisenberg exchange
$J$. Apart from the obvious increase in $T_c$ with $J'$ in
both cases, and the larger increase at higher electron density,
we note that even at a low density like $1.5\%$ a coupling
$J' = 2t$ (typical of the polaronic phase) almost doubles
the $T_c$! The doubling is of course because we have
taken a low value, $J=0.1t$. Superexchange couplings 
are typically a few percent of $t$, while electron-spin
couplings are $O(t)$, so even a dilute electron system
will affect the $T_c$ noticeably. Thankfully, it does not
change the nature of the ground state.

\begin{figure}[t]
\centerline{
\includegraphics[height=3.6cm,width=8.5cm]{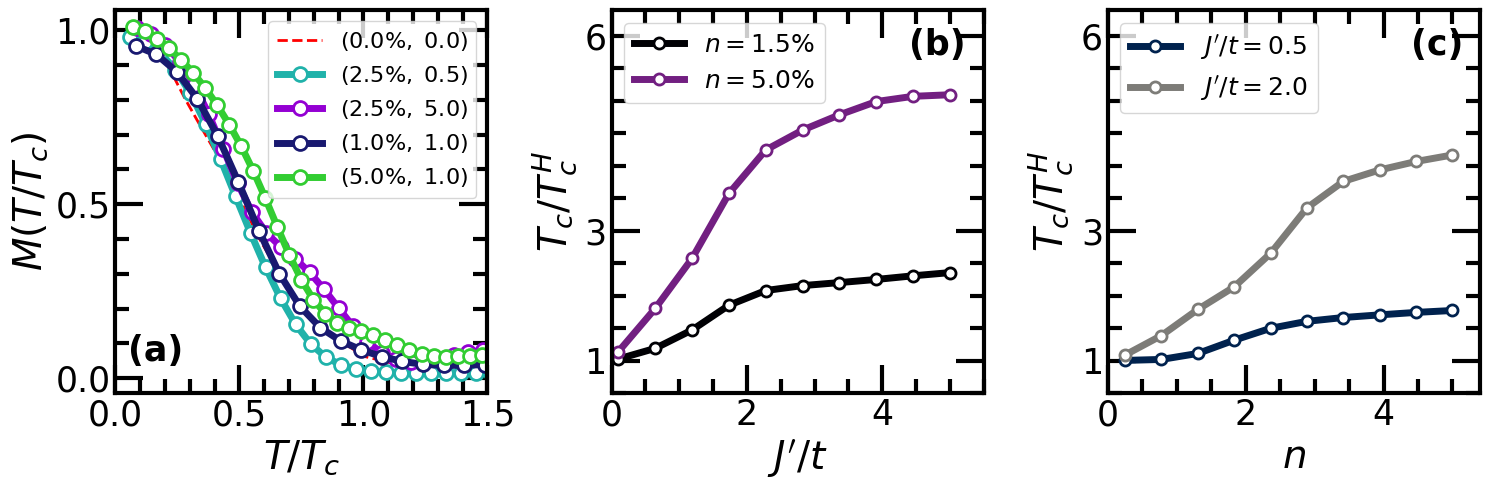}
}
\caption{Scaling of the magnetisation, and $T_c$ scales.
(a)~Plot of $m(T/T_c)$ to check that the overall behaviour
has a similar pattern independent of $n$ and $J'$. The 
approximate collapse will allow us to encode magnetic 
information only in terms of $T_c$. The legend indicates $(n, J'/t)$ values corresponding to each curve. 
(b)~$T_c(J')$ for two values of $n$.
(c)~$T_c(n)$ for two values of $J'$.
}
\end{figure}

Since we will scale all temperatures with respect to $T_c$
we wanted to check that the magnetisation profile has a 
reasonably common look when scaled by $T_c$, i.e,
$m(T, n, J') \approx m(T/T_c(n,J'))$. Fig.3(a) shows this
`collapse'. This is hardly surprising since the $m(T)$
is constrained for all $n$ and $J'$ by $m=1$ at $T=0$
and $m \rightarrow 0$ for $T \gg T_c$ (on a finite size)
and the magnetic model has the same symmetry throughout.
The plot shows data for a combination of $n$ and $J'$.

Fig.3(b) shows the increase of $T_c$ with respect to the
Heisenberg result at two values of $n$, as $J'$ is changed.
They have a small $J'^2$ window, hard to access on our
lattice size (where the smallest $J'$ is $0.5t$), and
both curves saturate for $J'/t \gtrsim 4$. The
`double exchange' limit, it seems, is quickly reached. The
actual enhancement of $T_c$ in the large $J'$ limit is
$\sim 2.5T_c^H$ at $n=1.5\%$ and $\sim 5 T_c^H$ at $n=5\%$.
Fig.3(c) shows the $n$ dependence of $T_c$ for two values
of $J'$.

\section{Resistivity}

Our main focus in this paper is the resistivity, and we
would like to explore the temperature dependence at 
various $n$ and $J'$. From that, we want to infer the
phase diagram of Fig.1.

It is best to think in terms of four scans of the $n-J'$
plane. The simplest would be $J'/t \lesssim 1$ where
the magnetic state is only weakly affected by the electrons
and the scattering of electrons from the spin background 
can be captured with lowest order perturbation theory in
$J'$.
Fig.4(a) shows the result for $J'=0.5t$ and four values
of $n$. In all cases $\rho(T)$ vanishes as expected as
$T \rightarrow 0$, where the magnetic state is fully 
polarised and there is no scattering, and tend to a $n$
dependent constant for $T \gg T_c$, where the electrons
scatter from uncorrelated spin disorder.
Since $J'/t$ is small one expects that the $\rho(T)$ would
be decided by a Drude form: $\sigma(T) \sim ne^2 \tau/m$,
i.e, $\rho(T) \propto \Gamma(T)/n$, where the
scattering rate $\Gamma = 1/\tau
\propto J'^2$ at weak coupling.
The $T$ dependence of $\Gamma$ was proposed by Fisher and
Langer long back, we will discuss this later. At this moment
we just note that the high $T$ values in Fig.4(a) 
indeed vary as $1/n$.
We have also checked that even at low $n$ the spatial
density $n_i$, in individual magnetic configurations,
remains essentially homogeneous at all $T$. There are 
no polaronic effects.

\begin{figure}[b]
\centerline{
\includegraphics[height=7cm,width=8.5cm]{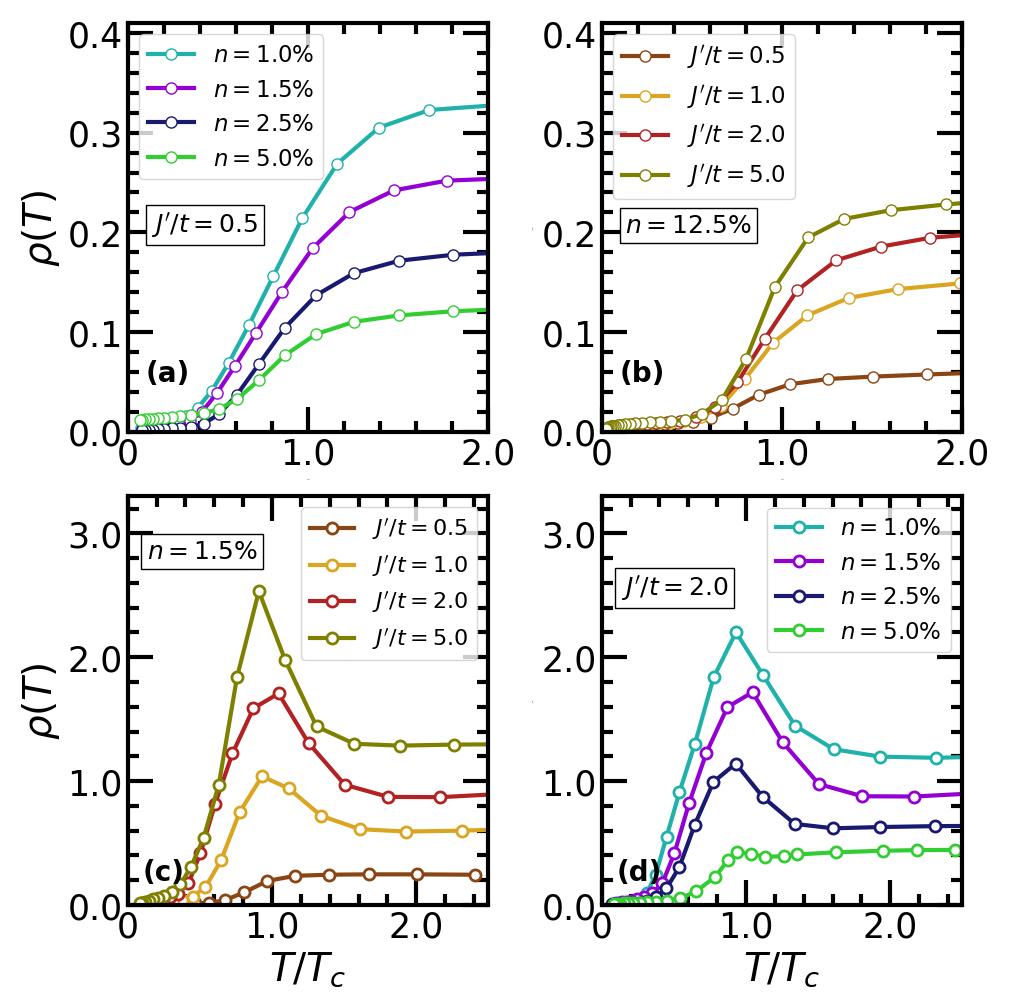}
}
\caption{
Resistivity $\rho(T)$. (a)~Density variation at
weak coupling, $J'=0.5t$. Note the monotonic $T$ dependence
of $\rho(T)$ and the increasing high $T$ saturation value
as $n$ is lowered. (b)~`High density' $n=12.5\%$ case for
varying $J'$. The temperature dependence shares a
similarity with (a). (c)~Increasing $J'$ at low density
$n=1.5\%$. While the weak coupling resistivity
is monotonic, for $J' \ge t$ the resistivity is distinctly
non monotonic, with a peak near $T_c$. The high $T$ value
also increases with $J'$. Essentially, at low $n$ one
enters an unusual transport regime with increasing $J'$.
Finally, (d)~Increasing density, starting from low $n$,
in the strong coupling regime. Here, the peak structure
weakens with increasing $n$ and almost vanishes for $n = 5\%$.
}
\end{figure}

We now want to check the $J'$ dependence, at reasonably
high density where we do not expect polarons. Fig.4(b)
shows results at $n=12.5\%$ for varying $J'$ - all the
way from $0.5t$ to $5t$. The $\rho(T)$ looks similar to
what we saw in panel (a), except the high $T$ values are
now functions of $J'$. Note that the high $T$ values 
{\it do not} increase as $J'^2$ at large $J'$. 
The next two panels, (c) and (d),
would bring in $T$ dependence quite unlike what we see
in (a) and (b).

Panel (c) shows the effect of increasing coupling at a
low density $n=1.5\%$. At $J'=0.5t$, the lowest curve,
the resistivity is monotonic. The three other curves,
for $J'=t$ and above, shows a clearly non monotonic
behaviour. There is an `excess resistivity', with
respect to a hypothetical monotonic background 
(see later), for $T$ around $T_c$.

Panel (d) shows density increase at $J'=2t$.
It has the $n=1.5\%$, $J'=2t$ data in common with
panel (c). If we decrease $n$ with respect to $n=1.5\%$
we find an enhanced peak in $\rho(T)$, and a higher
high temperature value. If we increase $n$ we
find a weakening of the peak and a move towards a
more `normal' resistivity.

\begin{figure}[b]
\centerline{
\includegraphics[height=7cm,width=8.5cm]{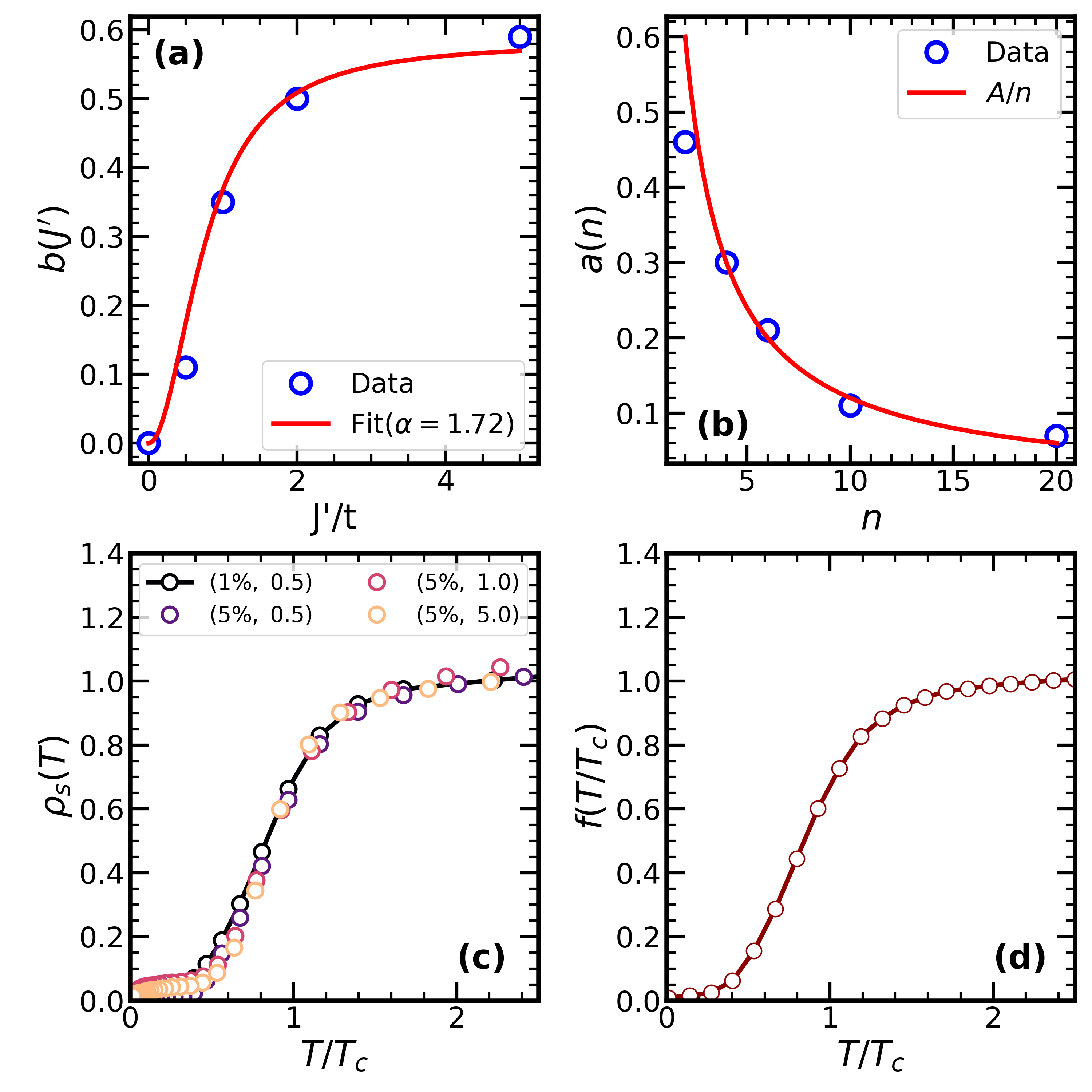}
}
\caption{Fitting functions for the resistivity $\rho(T)$.
(a)~The $J'$ dependence fitted to a form $J'^2/(1 + \alpha
J'^2)$ shows a reasonable fit with $\alpha = 1.72$ for all $n$.
(b)~A fit to the $n$ dependent prefactor to a form $A/n$
in the window $0-10\%$, gives
$A \sim 1.2$. (c)~The scaled
resistivity $\rho(T/T_c)/\rho_{\infty}$ for some combinations
of $n$ and $J'$. (d)~The best fit $f(T/T_c)$ to
the normalised $T$ dependence shown in (c).
}
\end{figure}
\begin{figure}[t]
\centerline{
\includegraphics[height=6.5cm,width=7.5cm]{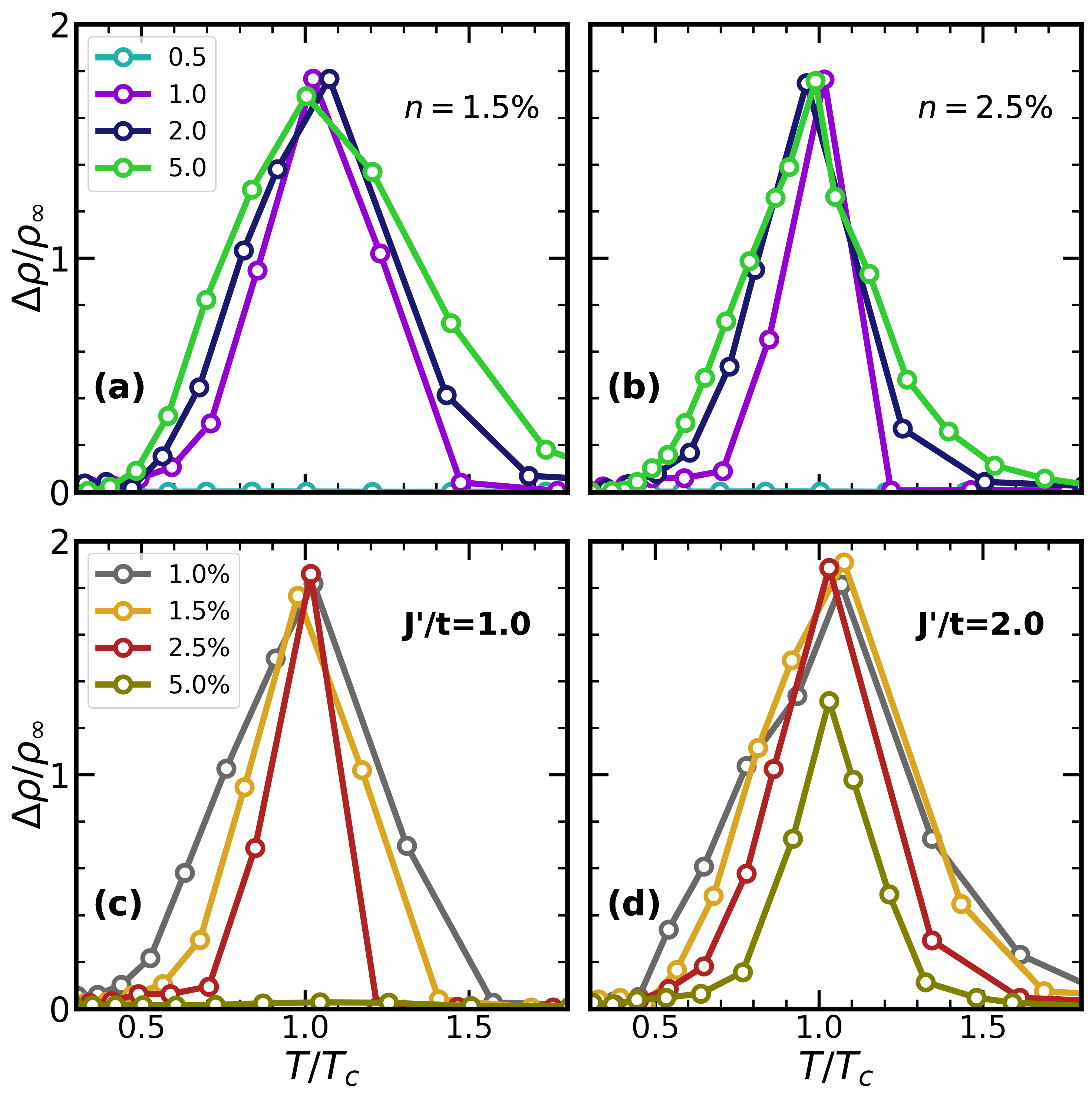}
}
\caption{The normalised `excess resistivity' defined by
$ \delta \rho(T/T_c) = 
(\rho(T/T_c) - \rho_{\infty} f(T/T_c))
/\rho_{\infty}$. (a)-(b) shows $J'$ variation at two $n$, while
(c)-(d) show $n$ variation at two $J'$. All these points are
in the polaronic window in Fig.1. Notice that the overall 
magnitude of $ \delta \rho$ is $\lesssim 1$, indicating that
the overall magnitude of $\rho(T)$ is set by $\rho_{\infty}$
- however deep one may be in the polaronic regime. While
the temperature window over which $\delta \rho$ is non zero
seems to be similar in all cases, $\approx 0.5T_c - 2T_c$,
the windows are somewhat larger at small $n$ and large $J'$.}
\end{figure}

Let us try to make some sense of the resistivity results.
While the weak coupling resistivity is expected to have
an overall scale varying as 
$(1/n)J'^2$, as $J'$ increases the quadratic
dependence on $J'$ would be increasingly inaccurate. The 
scattering rate would pick up contributions of
order $J'^4$, $J'^6$ from multiple scattering. 
Remarkably, we know the answer for $J'/t \rightarrow \infty$,
the `double exchange' limit, where $J'$ gets absorbed in
the chemical potential and one obtains a hopping disorder
model (see Discussion) whose overall scale is $t$  
\cite{ref_theo_other_montecarlo,ref_large_j1,ref_large_j2}.

Based on this, the weak coupling resistivity for $T \gg T_c$ 
would be $\propto (1/n)J'^2/t$, while at strong coupling 
the $T \gg T_c$ resistivity should be $\propto (1/n) t$.
The scattering rates could have a density dependence via
the density of states, but in 2D and low density the 
density of states is energy independent.
We can try an interpolative
form for the high temperature resistivity $\rho(T \gg T_c)
\equiv \rho_{\infty}$, given by:
$ \rho_{\infty} = (A/n) (J'^2/t)/[1 + \alpha (J'^2/t^2)]$.

For $J'/t \rightarrow 0$ this gives $\rho_{\infty} \sim 
({1 \over n}) J'^2/t$,  and for $J'/t \rightarrow \infty$
it gives  $\rho_{\infty} \sim  ({1 \over {n}}) t$,
satisfying both the limits. 
This raises three 
questions: (i)~does this fit our grid of $n-J'$ data?
(ii) after scaling $\rho(T)$
by $\rho_{\infty}$ we get a normalised $\rho$ that varies from
$0-1$ with $T$ - is this an universal function of $T/T_c$
and not dependent separately on $J'$ and $n$?
(iii)~what is the value of $\alpha$? 
If the answers to (i) and (ii) are in the affirmative 
then we have a complete characterisation of at least
the non polaronic window, and can examine the polaronic
regime with respect to this reference.

\begin{figure*}[t]
\centerline{
\includegraphics[height=5cm,width=12.5cm]{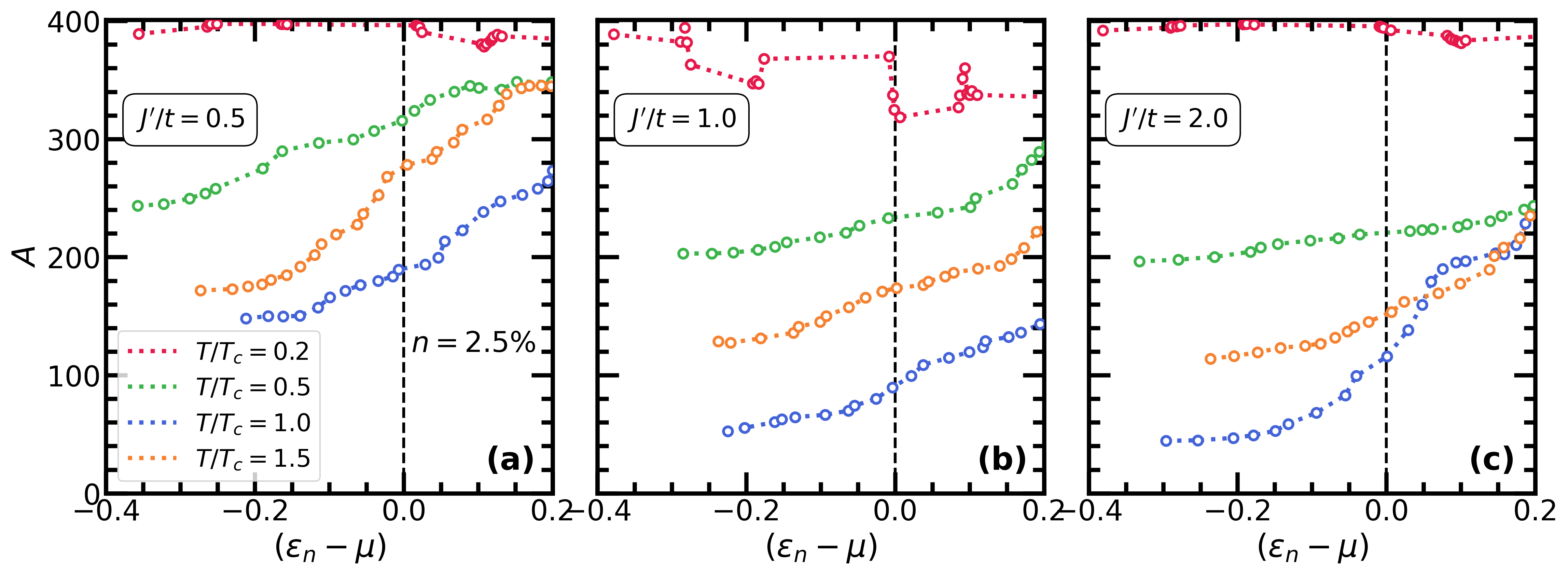} }
\centerline{
~~~~~~~~~~~~~~~~~
\includegraphics[height=4.6cm,width=13.5cm]{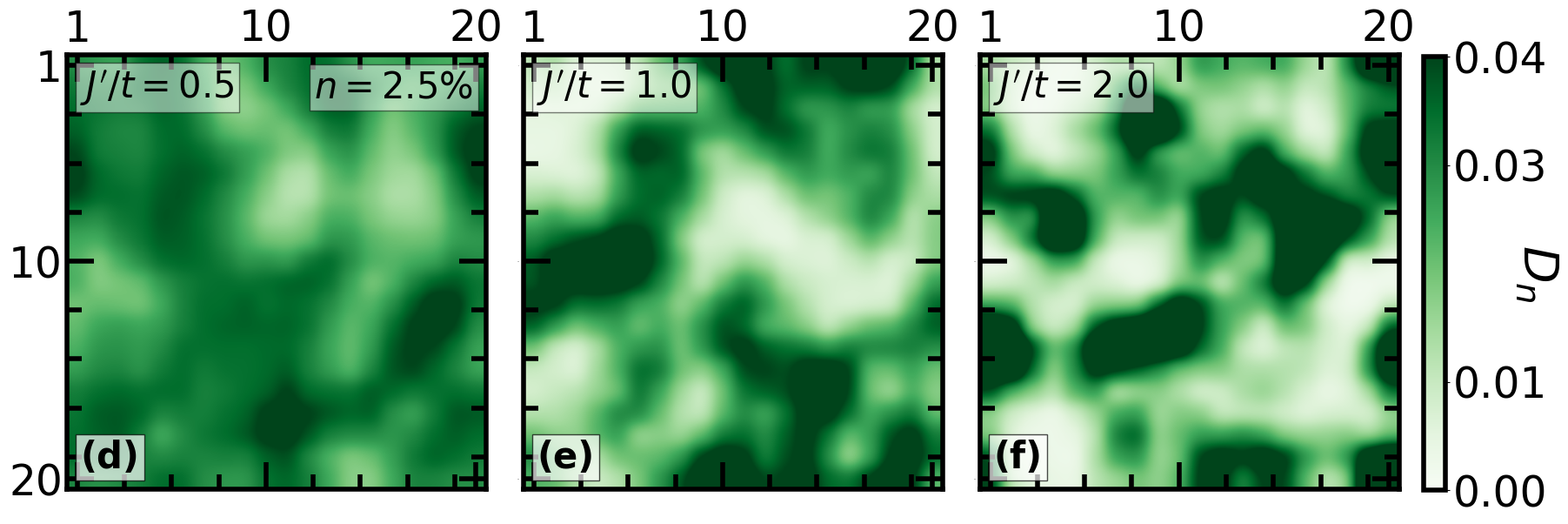} }
\caption{
The area $A$ associated with an eigenstate in 
typical magnetic configurations at different $T/T_c$, for
$n=2.5\%$ and $J'/t = 0.5,~ 1.0,~ 2.0$.
(a)~$A_n$ is $\sim$ system size at $T=0.2T_c$ for all $n$.
With $T \rightarrow T_c$ the $A_n$ for the occupied states
decrease quickly, but even at the lowest, they are $\sim 150$.
Beyond $T_c$ the $A_n$ rise again. (b)~Same density but $J'=t$.
Again $A_n \sim 400$ at low $T$, but at $T_c$ the occupied
states have $A_n$ ranging from $50-100$. The density field,
as we see later, has a textured pattern, but the localisation
near $T_c$ is still weak, $N_{el} A_{av} > N$, with $A_{av}
\sim 75$. For $T$ increasing beyond $T_c$ the $A_n$ increase.
(c)~Now $J'=2t$. In the same trend as in (b), we now have 
even smaller values of $A_n$ for the occupied states at $T_c$,
but still $N_{el} A_{av} > N$ - no well separated polarons. 
At $n=2.5\%$ we will need $A_{av} \sim 40$ for well localised
polarons to exist.
}
\end{figure*}

We try out the fitting.
First, at a fixed density, $n=12.5\%$ we fit $\rho_{\infty}$
to the form $ (J'^2/t)/(1 + \alpha (J'^2/t^2))$. The
quality of the fit is shown in Fig.5(a), and gives $\alpha = 1.72$.
When we try this at other $n$, putting in the prefactor $1/n$
we find the $\alpha$ to be independent of $n$. On the whole
$\rho_{\infty}$ can be fitted to the form:
$$\rho_{\infty} 
= 
({A \over n}) {{J'^2/t}
\over {( 1 + \alpha (J'^2/t^2))}}
$$
with $A \approx 1.2$, $\alpha \approx 1.72$.

What about the actual temperature dependence? 
We plot the scaled resistivity $
f(T)  = \rho(T)/\rho_{\infty}$
for several $n$ and $J'$ combinations in Fig.5(d) 
with respect to $T/T_c$. Given the finite size issues
and uncertainty in $T_c$ determination, we think one
can reasonably argue for ``collapse'' to a single
characteristic. It is possible that detailed calculations
on the Heisenberg-Kondo model in 3D in the tractable
$J'/t \ll 1$ and $J'/ \gg 1$ limits will reveal 
small differences between the scaled resistivity, but
the larger characterisation of the metallic phase
should survive. So $\rho(T, J, n) = \rho_{\infty}(J, n) 
f(T/T_c)$.

What if we test out this function in the polaronic regime,
now that we know it's $n$ and $J'$ dependence?
The fit works
well at low and high $T$ but we obtain an 
excess resistivity around $T_c$, defined by
$$
\delta \rho(T/T_c) = {{\rho(T/T_c) - \rho_{\infty} f(T/T_c)} 
\over {\rho_{\infty}}}
$$
This is shown in Fig.6. Panels (a)-(b) show the $J'$ dependence
at $n=1.5\%$ and $2.5\%$, respectively, while (c)-(d)
show the $n$ dependence at $J'=t$ and $J'=2t$.
The first observation is that the peak in $\delta \rho$ is
always in the range $1.8-2.0$, which means it scales with
$\rho_{\infty}$ at that $n$ and $J'$. 
The peak occurs around
$T_c$, and the function is non zero in the interval $\sim 0.5T_c
- 2T_c$.
Looked at closely the temperature window does have some
variation with $n$ and $J'$, but for the moment let us
focus on addressing the overall structure of $\delta \rho$.
Does it arise from 
\begin{itemize}
\item
Localisation of states at the chemical potential as $T$
increases towards $T_c$ and progressive delocalisation
thereafter? Or, is it related to 
\item
An enhancement in the scattering rate of delocalised 
states at the chemical potential as $T$ increases towards 
$T_c$, and its decrease thereafter?
\end{itemize}
This requires us to examine the eigenstates 
in typical magnetic configurations at various $T$ for 
different combinations of $n$ and $J'$.

\section{Eigenstates and mobility edge}

The inverse participation ratio (IPR) is the standard measure of
localisation of single particle states. We use the inverse of
this object as a measure of the area covered by an eigenstate.
Thus $A_n = [\sum_i \vert \psi^n_i  \vert^4]^{-1}$,
where we assume a normalised state.
For a plane wave, we would have $\psi^n_i 
= 1/\sqrt{N}$, so $A_n$ would be $N$, the number
of lattice sites. Correspondingly, a state distributed 
uniformly over $N_p$ sites will have $A = N_p$. 

Fig.7 shows $A_n$ for electronic eigenstates $\psi_n$ 
from typical equilibrium spin configurations at four temperatures
from $0.2T_c$ to $1.5T_c$. Panel (a) shows $A_n$ at $n=2.5\%$ 
and $J'=0.5t$. Negative values of $\epsilon_n - \mu$ 
correspond to the occupied states
(modulo Fermi function effects). At $T=0.2T_c$ where the magnetic
state is essentially polarised the electrons see very little 
disorder, and $A \approx 400$ for all  $\epsilon_n$ on our $20 \times
20$ lattice. With $T$ increasing towards $T_c$, the $A_n$ associated
with the occupied states drop, reaching a minimum $\sim 150$ sites
at $T_c$. The average $A_{av}$ of $A_n$ over the occupied states
at this $T$ is $\sim 170$. 
This is too large for isolated 
polarons in a $N_{el}=10$
system, since $N_{el} \times A_{av} \sim 1700 \gg 400$, the system 
size. We have also confirmed that the spatial density remains
essentially uniform at $T_c$. With $T $ increasing beyond $T_c$ 
the
$A_n$ increases. So, no polarons at this $(n,~J')$ point, 
at any $T$, consistent with the monotonic $\rho(T)$.

Panel (b) shows results at the same density but $J'=t$. All the
$A_n$ traces are systematically lower than the corresponding $J'=0.5t$
case, but even here $A_{av}$ at $T_c$ is $\sim 75$, with $N_{el}
A_{av} > N$. We see an increasing tendency towards localisation
but no well separated polarons yet. Panel (c), at $J'=2t$, shows
a similar trend, with $A_{av} \sim 75$ at $T_c$, not small enough
for isolated polarons. What we see in (b) and (c) are reasonably
compact low energy states but as $\epsilon_n \rightarrow \mu$
there is a rapid increase in $A_n$. The states near $\mu$,
which will decide the d.c conductivity can be called weakly
bound or strongly scattered at this system size. 

\begin{figure}[b]
\centerline{
\includegraphics[height=6cm,width=6.0cm]{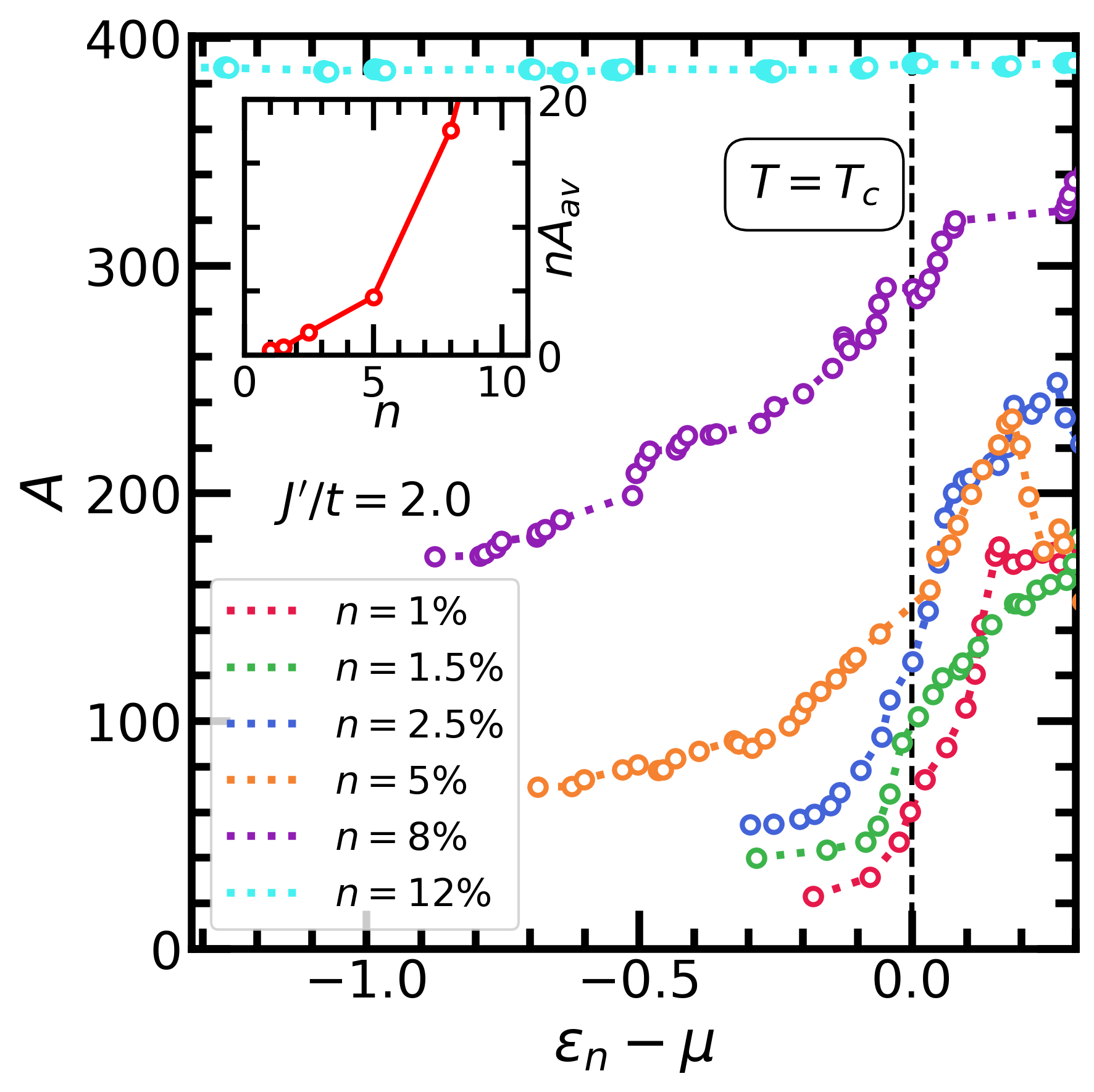}
}
\caption{The area $A_n$ for several $n$ at $J'=2t$ and $T_c$.
At the highest density, $n=12\%$, all states remain delocalised.
There is a suppression in the $A_n$ for the occupied states
as $n$ decreases to $8\%$ and $5\%$, but the $A_{av}$ for the
occupied states is still too large. For the lowest two $n$
however the occupied states are sufficiently compact
for the system to be characterised as polaronic.
}
\end{figure}

\begin{figure}[t]
\centerline{
\includegraphics[height=5cm,width=8.6cm]{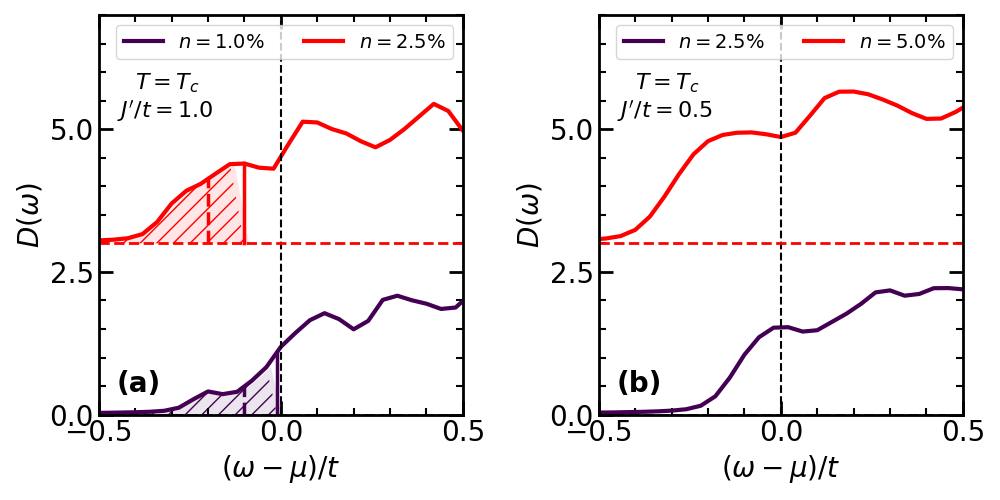}
}
\caption{The density of states, showing the `mobility edge'
separating the localised (shaded) states from the delocalised
ones. (a)~$J'=t$ and $n=1\%$ and $2.5\%$, and $T=T_c$. At the
higher $n$, the mobility edge remains below $\mu$, while at
the lower $n$, it almost coincides with $\mu$. All this is
assuming $A \le 60$ as a localised state. 
(b)~Same as (a) but now with $J'=0.5t$. There are no 
localised states and the mobility edge is at the band
bottom.
}    
\end{figure}

We now focus on $T_c$ and show a wider density variation at $J'=2t$.
Fig.8 shows that at high density, $n=12\%$, we have 
$A_n \sim 400$ even at $T_c$
The $A_n$ profile gets
lowered systematically as $n$ decreases. However, it is only for
the lowest two plots, $n=1\%$ and $1.5\%$, that $N_{el} A_{av}
\lesssim N$. 
When this product is $\lesssim 1$ we have a well localised
polaronic phase.

Since the polarons are localised states, in contrast to the higher
energy delocalised states, we now go through the exercise of
identifying a `mobility edge' $\epsilon_c$ in the density of
states, at a given $T$. This is a risky venture since locating
$\epsilon_c$ in a standard disordered system involves
systematic increase in system size and locating which
states have IPR $O(1)$ (localised) versus IPR $O(1/N)$ 
(delocalised). Given the computational cost of our
basic calculation, we cannot do a significant size variation
and arbitrarily set $A_n = 60$ as cutoff for a localised
state. 
We also checked what happens when we set a cutoff of 40 -
the dashed line in Fig.9 shows this.

Fig.9 shows the mobility edge, using this criterion, for two
densities, $n=1\%$ and $n=2.5\%$ and $J'=1$, at $T=T_c$.
Panel (a) shows that $\epsilon_c$ remains below, but close to
$\mu$ at $n=2.5\%$ while at  $n=1\%$ 
it actually coincides with $\mu$.
Fig.10 will show how $\epsilon_c - \mu$ 
varies with temperature
in a few cases. 
Panel (b) shows the DOS for the same two densities
but now at $J'=0.5t$ and $T=T_c$. There are no
localised states in this case and the mobility
edge is basically at the lower edge of the band.

\begin{figure}[b]
\centerline{
\includegraphics[height=6cm,width=6.0cm]{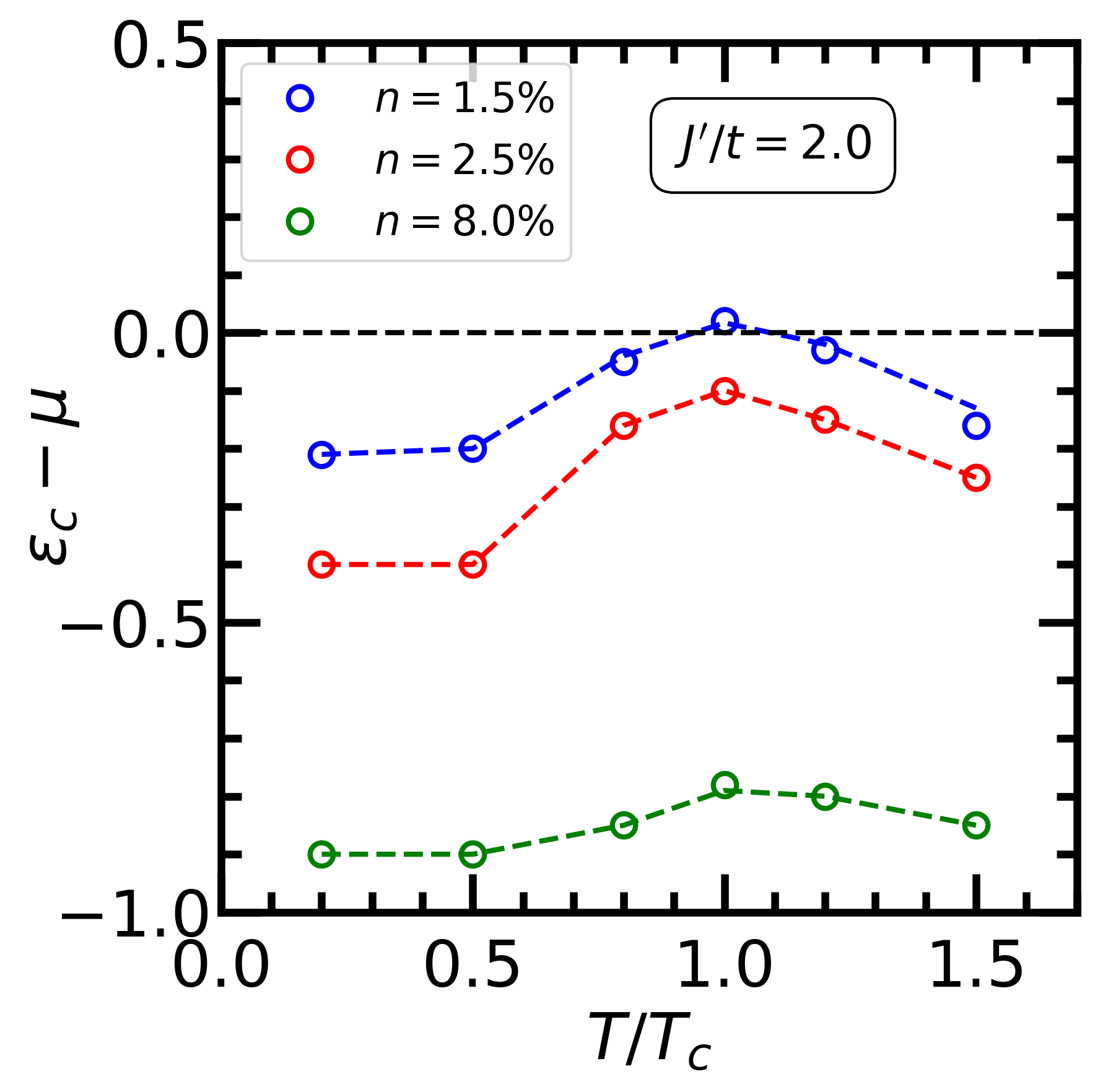}
}
\caption{The temperature dependence of the mobility edge,
with respect to the chemical potential, at $J'=2t$ and
three values of $n$. In all cases $\epsilon_c$ starts
at the band bottom at $T=0$ and essentially stays there
for $T \lesssim 0.5T_c$. After that, it rises, has a maximum
near $T_c$, and drops again. For $n=1\%$, and with our
defnition of localised states, the mobility edge does
cross $\mu$ for $n=1\%$, but for the higher densities
it approaches $\mu$ but never crosses.
}
\end{figure}

With all the uncertainty about the accurate location of
$\epsilon_c$ in the DOS, we can still track its $T$
dependence using the criteria chosen.
Fig.10 shows $\epsilon_c(T) - \mu(T)$ for densities
$n=1,~2.5,~5\%$ at $J'=2t$. In all three
cases  $\epsilon_c - \mu$ is `flat' for $T < 0.5T_c$,
remaining pinned at the lower edge of the band.
Beyond $T \sim 0.5T_c$ it starts rising, has a peak
near $T_c$, and drops back again. At the two higher
densities $\epsilon_c$ approaches the chemical
potential and then recedes at higher $T$, while at
$n=1\%$ it actually crosses $\mu$ and falls below
again as $T$ crosses $\sim 1.5T_c$.

This result is important in creating an understanding
of the excess resistivity, with respect to perturbative
scattering, that one sees in the polaronic regime.
It seems in general that a `mobility gap', $\epsilon_c 
- \mu > 0$,
over a certain temperature window around $T_c$, is not
the generic scenario behind the non monotonic resistivity.
States near $\mu$ have a spatial extent which are a fair
fraction of system size. The non monotonicity in the area
$A(\mu)$ with $T$ and now the  non monotonicity of 
$\epsilon_c - \mu$ correlates with the non monotonicity of
$\rho(T)$, but we want a more quantitative connection.
With that in mind, we opted to compute the electron
spectral function $A_{\sigma}({\bf k}, \omega)$, 
for the small ${\bf k}$ values in the low density window,
and check the scattering rate from its width.

\section{Single particle scattering rate}

The conductivity is determined by the current-current correlation
in equilibrium magnetic configurations, and then thermally
averaged. Within our model of electrons coupled to classical
spins this can be written in terms of the `unaveraged'
Green's functions $G_{\sigma \sigma'}({\bf k}, {\bf k}', \omega)$.
The spin and momentum off-diagonal components of $G$ exist
since the electrons can undergo spin-flip scattering from
the magnetic background.

\begin{figure}[b]
\centerline{
\includegraphics[height=4.5cm,width=8.5cm]{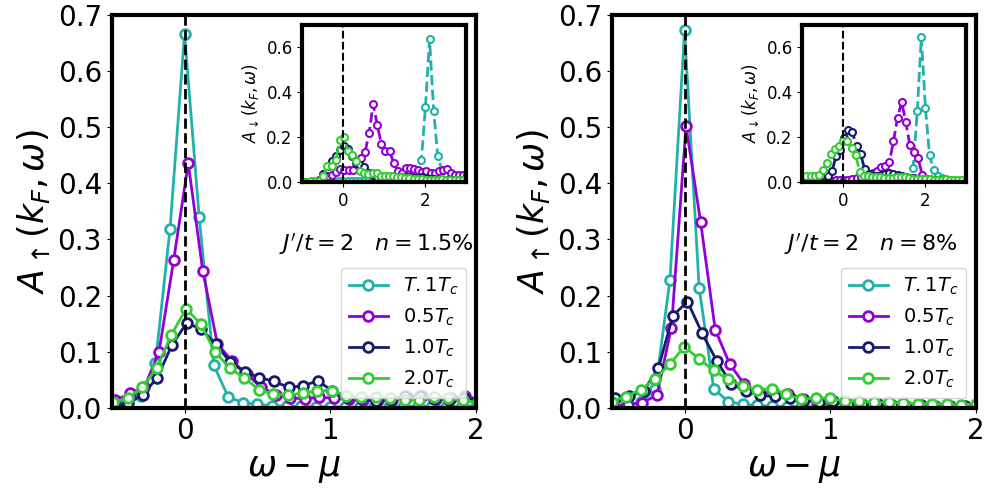}
}
\caption{Single particle spectral function $A_{\sigma}(k_F,
\omega)$: We compute the spectral function using the exact
eigenstates and eigenvalues in individual equilibrium spin
configurations and then average thermally. (a)~The main
panel shows
$A_{\sigma}(k_F, \omega)$ in the polaronic regime, with a
non monotonic behaviour as a function of $T$. (b)~Shows the
result at the same $J'$ but larger $n$, where no polarons exist,
and the spectra broaden monotonically with increasing $T$.
The insets show the spectra in the minority spin channel.
}
\end{figure}
\begin{figure}[t]
\centerline{
\includegraphics[height=5.1cm,width=5.5cm]{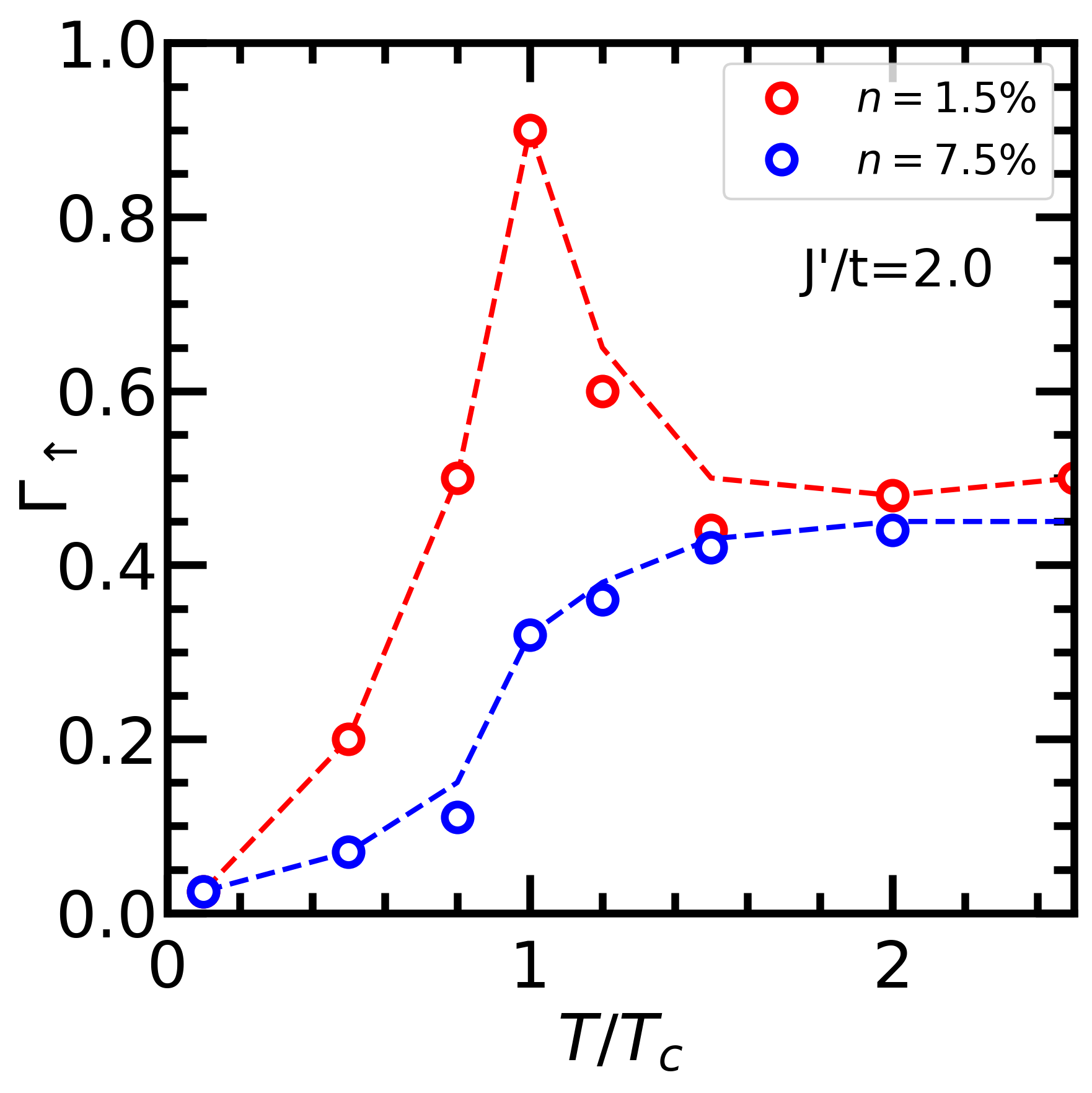}
}
\caption{Scattering rate in the polaronic and non polaronic
regimes. We extract the half width of $A_{\uparrow}(k_F, \omega)$
and show it's temperature dependence in the polaronic $n=1.5\%$
and non polaronic $n=7.5\%$ cases. The prominent peak in the
polaronic case correlates with the behaviour of the resistivity.
}
\end{figure}

To get an estimate of the scattering rate we chose to compute
the spectral function 
$$
A_{\sigma}({\bf k}, \omega)
= -({1 \over \pi}) 
Im \big[G_{\sigma \sigma}({\bf k}, {\bf k}, \omega) \big]
$$
This can be related to the overlap of momentum eigenstates 
$\vert {\bf k}, \sigma \rangle = c^{\dagger}_{{\bf k}, \sigma}
\vert 0 \rangle$ with the exact eigenstate $\vert n \rangle 
= a^{\dagger}_n \vert 0 \rangle$
of the electron in the magnetic background. 
Denoting $\langle n  \vert {\bf k}, \sigma \rangle  
= u^n_{{\bf k}, \sigma}$,
we have 
$$A_{\sigma}({\bf k}, \omega)
 = \sum_n \vert  u^n_{{\bf k}, \sigma} \vert^2 
\delta(\omega - (\epsilon_n - \mu))
$$
We know the eigenvalues and eigenfunctions in each equilibrium
spin background and so can calculate $A$, and average it over
thermal configurations at a given $T$.

The expression we had suggested earlier to model the 
background resistivity is based on leading order 
perturbation theory, and would work if the chemical
potential is far from a mobility edge. However if
$\epsilon_c$ approached $\mu$ the scattering rate 
$\Gamma_{\sigma}$ would exceed the background value.
Rather than attempt a complicated approximation for
the $\Gamma$ we chose to extract it directly from the
width of the spectral function at $k \sim k_F$ (at the
low $n$ that we use the FS is circular). 

Fig.11(a) shows 
$A_{\sigma}({\bf k}, \omega)$ 
in the polaronic phase, $n=1.5\%$ and
$J'=2t$, at four temperatures. The main panel shows the
spectra for spin up (majority) while the inset shows the
result for spin down. At $T \ll T_c$ transport will be
dominated by the up spin channel, while for $T > T_c$
the up and down spin spectra would be the same. The
broadening in the feature around $\omega = \mu$, with
increasing $T$ is obvious. While the width itself 
requires a numerical estimate (which we do next) note
that the height $A_{\uparrow}(0)$ first falls with
increasing $T$ and then weakly rises past $T_c$.
Panel (b) shows the result at a non polaronic point,
again $J'=2t$ but at $n = 7.5\%$. Here the amplitude
$A_{\uparrow}(0)$ decreaes monotonically with $T$ and
the width visibly increases.

Fig.12 shows the broadening $\Gamma_{\uparrow}$, 
indicative of the scattering rate, as a function of 
temperature.
The red symbols correspond to the polaronic parameter
point, the blue to the non polaronic.
$\Gamma_{\uparrow}$ in the non polaronic case is 
monotonically increasing, while in the polaronic case
it has a peak near $T_c$ and falls to a value
characteristic of the $J'$ value. Compare this with
the $T$ dependence of $\epsilon_c - \mu$ in Fig.10,
where $\epsilon_c$ approaches $\mu$ and again draws
away as $T$ goes past $T_c$. 

This shows that a polaronic phase can have a spontaneous density
inhomogeneity but the states near $\mu$ need not be localised
for an anomaly to be seen in the scattering rate and resistivity.
This is more subtle than a `strong localisation' scenario (which
anyway does not happen without a pinning disorder. where a mobility
gap appears in the density of states and one gets activated
transport in the paramagnetic phase.

\section{Conclusion}

In this paper, we have established the regime in electron-spin
coupling $J'$ and density $n$ over which magnetic polarons can exist 
in the ferromagnetic Heisenberg-Kondo lattice model. We find that 
polarons exist only above a critical coupling $J' \sim 0.7t$, and no 
polarons are possible for density $n \gtrsim 5\%$, however large the 
coupling. The polaronic phase exists when $J' > J'_c(n)$. The 
temperature window around $T_c$ over which polaronic effects are seen 
shrinks as $J'$ reduces towards $J'_c(n)$, 
and saturates for $J' \gg J'_c(n)$.
We find that in the non polaronic regime the resistivity $\rho(T, J', n)$ 
is reasonably described by a form $(1/n) J'^2/(1 + J'^2) f(T/T_c)$, 
all the way from weak to strong coupling, where the function $f$ rises
from zero at $T=0$ and saturates for $T \gg T_c$. The polaronic phase, by 
contrast, shows an `excess resistivity' beyond this background, with a 
peak near $T_c$. To trace its origin, we examined the inverse participation 
ratio (IPR) of electronic states in the equilibrium magnetic backgrounds 
at different temperatures. We find that in the non polaronic regime all 
states are extended at all temperatures, while for a parameter point in
the polaronic window a fraction of states below the chemical potential 
get localised as temperature rises towards $T_c$. This fraction rises
as $T \rightarrow T_c$ and falls again as $T$ increases beyond $T_c$.
Typically, except at very low $n$, the mobility edge only approaches
the chemical potential as $T \rightarrow T_c$ but does not cross.
In this situation, the transport would be determined by a scattering 
rate that is enhanced beyond the naive perturbative value for
$T \rightarrow T_c$ and falls thereafter. To check this possibility,
we calculated the electronic spectral function $A_{\sigma}({\bf k},
\omega)$ and probed the quasiparticle scattering rate as a function
of temperature. It indeed shows a non monotonicity in the polaronic
regime and mimics the monotonic resistivity in the non polaronic
window. 
On the whole, we find that a low density 
strong coupling electron-spin
system can display spontaneous density inhomogeneities
but not strong localisation. The anomalies in resistivity
arise from enhanced scattering due to proximity to
a mobility edge, rather than genuine localisation and a gap
in the spectrum.

{\it Acknowledgment:} We acknowledge use of the HPC clusters
at HRI.


\end{document}